\documentclass[a4paper, amsfonts, amssymb, amsmath, reprint, showkeys, nofootinbib, twoside, superscriptaddress]{revtex4-1}
\usepackage[english]{babel}
\usepackage[utf8]{inputenc}
\usepackage[colorinlistoftodos, color=green!40, prependcaption]{todonotes}
\usepackage{amsthm}
\usepackage{mathtools}
\usepackage{physics}
\usepackage{xcolor}
\usepackage{graphicx}
\usepackage[left=18mm,right=18mm,top=35mm,columnsep=15pt]{geometry} 
\usepackage{adjustbox}
\usepackage{placeins}
\usepackage[T1]{fontenc}
\usepackage{lipsum}
\usepackage{csquotes}
\usepackage{siunitx}
\usepackage{caption}
\usepackage{subcaption}
\usepackage[pdftex, pdftitle={Article}, pdfauthor={Author}, colorlinks=true]{hyperref}
\usepackage{multirow}
\begin{document}
\title{Decoherence rate expression due to air molecule scattering in spatial qubits}

\author{Martine Schut}
    \affiliation{Van Swinderen Institute for Particle Physics and Gravity, University of Groningen, 9747AG Groningen, the Netherlands }
    \affiliation{Bernoulli Institute for Mathematics, Computer Science and Artificial Intelligence, University of Groningen, 9747 AG Groningen, the Netherlands \vspace{1mm}}

\author{Patrick Andriolo}
    \affiliation{{Physics Institute, University of São Paulo, Rua do Matão 1371, São Paulo, Brazil}}
    
\author{Marko Toro\v{s} }
    \affiliation{Faculty of Mathematics and Physics, University of Ljubljana, Jadranska 19, SI-1000 Ljubljana, Slovenia}

\author{Sougato Bose}
    \affiliation{Department of Physics and Astronomy, University College London, London WC1E 6BT, United Kingdom}

\author{Anupam Mazumdar }
    \affiliation{Van Swinderen Institute for Particle Physics and Gravity, University of Groningen, 9747AG Groningen, the Netherlands }


\begin{abstract}
We provide an expression for the decoherence rate in spatial superpositions due to scattering/collision with air molecules which is independent of the wavelength of the air molecules.
This result reproduces the short- and long-wavelength limits known in the literature. 
We compare the decoherence rate with several existing interpolations in the literature and evaluate the decoherence rate and experimental parameters when creating macroscopic quantum spatial superpositions (i.e., in micron-size spheres).
The interpolation regime is relevant in, e.g., matter-wave interferometry, where one might switch between the wavelength limits during the experimental protocol.
Finally, we consider the decoherence rate's time dependence while creating and closing the spatial superposition in an interferometer setup.
\end{abstract}


\maketitle

\section{Introduction}\label{sec:intro}

Matter-wave interferometers are at the heart of quantum mechanics, which follows the rules of wave-particle duality~\cite{de1923waves}.
It is also the basis for the notion of a quantum spatial superposition of matter~\cite{Schrodinger:1935zz}
, and serves as a building block for the quantum entanglement features that any classical theory cannot mimic~\cite{Horodecki:2009zz}. 
Various constructions of matter-wave interferometers are now under consideration to find a the Earth's gravitationally-induced phase, such as with the help of atoms ~\cite{asenbaum2017phase,fixler2007atom,Kovachi_2015} and cold neutrons~\cite{colella1975observation,nesvizhevsky2002quantum}.
Additionally, they have been proposed to use in tests for physics beyond the Standard Model~\cite{Barker:2022mdz,Elahi:2023ozf}, modified gravity~\cite{Marshman:2019sne,Vinckers:2023grv,Chakraborty:2023kel}, gravitational waves~\cite{Marshman:2018upe}, the quantum equivalence principle~\cite{Bose:2022czr}, quantum metrology/sensors~\cite{Wu:2022rdv, Wu:2024bzd,Toros:2020dbf}, and the quantum nature of gravity (the quantum gravity-induced entanglement of masses (QGEM) experiment)~\cite{Bose:2017nin,ICTS}, see also~\cite{Marletto:2017kzi}. 
The QGEM protocol motivated entanglement-based tests of the relativistic modifications to quantum electrodynamics~\cite{Toros:2024ozf} and quantum gravity~\cite{Toros:2024ozu}.

One of the concepts essential in any matter-wave interferometer is the notion of decoherence~\cite{Zeh:1970zz,
Joos:1984uk,Zeh:1999fs,Kiefer:1992cn,Zurek:1981xq,schlosshauer,Hornberger:2003,Hornmerger:2022,Romero_Isart_2011,RomeroIsart2011LargeQS,Schlosshauer_2019}. Any quantum interaction with the matter-wave interferometer will lead to decohere the quantum system. 
Additionally, dephasing can occur in the interferometer.
Decoherence and dephasing in specific experiments typically lead to loss of contrast in the visibility of the matter-wave interferometer.
While both cause coherence loss, decoherence is a fundamental loss of coherence, while dephasing can be mediated by better control of the system.
In nanosphere interferometers, the dominant coherence loss interaction seems to be that of the short and long-range quantum electrodynamics~\cite{vandeKamp:2020rqh,Kilian:2022kgm,Fragolino:2023agd,Schut:2023tce}.
The loss of coherence can have many different sources, such as a fluctuating magnetic field, fluctuations in the current~\cite{Zhou:2022jug,Zhou:2022epb,Zhou:2022frl,Zhou:2024pdl}, noise due to external jitters~\cite{Toros:2020dbf,Wu:2024bzd}, internal decoherence due to phonons~\cite{Henkel_2022,henkel2024limitspatialquantumsuperpositions,Xiang:2024zol}, and spin-induced fluctuations which is specific to a Stern-Gerlach type interferometer~\cite{Japha2022QuantumUL,Zhou:2024pdl}, and gravitational induced loss of coherence, see~\cite{Belenchia:2018szb,Toros:2020krn,Danielson:2022tdw}.

Typically, the decoherence rate can be computed by studying the scattering cross-section, which often scales with the radius of the test mass. As a result, mesoscopic test masses (i.e., nano/microspheres) face high decoherence rates by their large cross-sectional area. This leads to a fundamental loss of quantum coherence in the system and, hence, loss of visibility/contrast in the matter-wave interferometer. 
Such loss of contrast is a primary concern for experiments that propose testing fundamental physics, such as the quantum nature of gravity with neutral nanospheres in a Stern-Gerlach interferometer~\cite{Bose:2017nin,Margalit:2020qcy}.
These setups propose using the embedded spin in a neutral microsphere of nanodiamonds to create a spatial superposition via a Stern-Gerlach interferometer. 

To create a spatial superposition for a massive object, one needs to pursue the purest quantum state of a particle, which is very challenging. Typically, these constraints surmount to cooling of the centre of mass motion (C.O.M) of the particle~\cite{Deli__2020,Piotrowski_2023} and the cooling of the interior degrees of freedom, such as phonons~\cite{Ju:2024nli}. The latter also means that the objects should not emit or absorb photons, e.g., blackbody radiation. Creating mesoscopic superpositions with long coherence times at low temperatures is especially prohibited by the scattering of air molecules.
Since the wavelength of the air molecules is small compared to the size of the spatial superpositions proposed in fundamental physics experiments, scattering with air molecules removes a lot of quantum information from the test masses. Since the size of the spatial superposition is time-dependent, this, of course, depends on the dynamics of the superposition.

In the literature, decoherence has been studied quite rigorously using the scattering model~\cite{schlosshauer,Romero_Isart_2011,Hornberger:2003,Hornberger,Arndt_2014,Hornberger}, and has also been studied from a field theory approach~\cite{Fragolino:2023agd,Kilian:2022kgm}. The standard practice for air molecule scattering, photon scattering, and blackbody emission and absorption is to solve the decoherence rate in the limit where the environmental particle's wavelength is big or small compared to the size of the spatial superposition, see~\cite{Romero_Isart_2011,Biswas:2022qto,carlesso2016decoherence}. 

This paper will aim to find a solution for the decoherence rate of air molecule scattering that does not use the approximation that the wavelength of the external particle is either very long or very short compared to the size of the superposition. 
Our resulting expression interpolates the two regimes of long and short wavelengths for air molecule scattering. 
The interpolating regime is important in matter-wave interferometry, where during the creation and annihilation of the spatial superposition, one might switch from the long-wavelength limit to the short-wavelength limit.
In this case, knowing the interpolation is key to precisely estimating the decoherence.
Notably, we have also considered the time dependence of the decoherence rate, i.e. spatio-temporal decoherence of the matter-wave interferometer during the creation and closing of the interferometric paths.
In section~\ref{sec:introduction}, we begin with a brief recap on spatial decoherence. 
In section~\ref{sec:air-molecules}, we provide an expression for the decoherence of the air molecule scattering. 
There, we also compare our results with the existing results in the literature. 
We apply our results to analyze the expected decoherence rate in mesoscopic spatial superpositions, resulting in the table~\ref{table:parameters} which shows the experimental parameters in such experiments. 
Considering a specific time-dependent path of the spatial superpositions, we also discuss the dynamical aspects of decoherence in section~\ref{subsec:time-dependence}.

\section{Recap of collisional decoherence}\label{sec:introduction}
This paper focuses on collisional decoherence, also called the scattering model of decoherence; for a review, see, e.g. refs.~\cite{Hornberger:2003,schlosshauer,Vacchini_2009,Schlosshauer_2019}. 
In the case of spatial decoherence, we can denote the scattering wavefunction $\phi(x)$ in the position basis $\ket{\boldsymbol{x}}$ and the state of the environmental particle as $\ket{\chi}$. 
Assuming that the test mass is much more massive than the environment's mass, meaning that the C.O.M of the mesoscopic object is not disturbed, the generic scattering of an ambient particle with the test mass results in an out-state of the test mass entangled with the out-state of the ambient particle~\cite{schlosshauer,Hornberger:2003,joos2013decoherence,Vacchini_2009}
\begin{align}
    \int \dd[3]{\boldsymbol{x}} \phi(\boldsymbol{x}) \ket{\boldsymbol{x}} \ket{\chi} 
    &\xrightarrow{t} 
    \int \dd[3]{\boldsymbol{x}} \phi(\boldsymbol{x})  \ket{\boldsymbol{x}} \ket{\chi_{\boldsymbol{x}}} \, . \label{eq:wf}
\end{align}
The state on the left-hand side in eq.~\eqref{eq:wf} above is separable, i.e. not entangled. After scattering (indicated by the time-evolution with the arrow), the environmental particle becomes entangled with the test mass; the wavefunction is non-separable because the scattering depends on the C.O.M. position of the test mass. Since the environmental particle's information is lost, this is classified as decoherence rather than entanglement.
The scattering between the test mass and environmental particle is characterized by the scattering matrix $\hat{S}$, with in eq.~\eqref{eq:wf}: $\ket{\chi_{\boldsymbol{x}}}=\hat{S}_{\boldsymbol{x}} \ket{\chi} $.
Based on the wavefunction in eq.~\eqref{eq:wf}, one can find the density matrix operator $\rho$, a tool often used to study the dynamics of quantum systems.
The dynamics of the system's density matrix ($\rho_\mathcal{S}$) due to the environmental scattering are then obtained by tracing out the environment's ($\mathcal{E}$) degrees of freedom, $\rho_\mathcal{S} = \Tr_\mathcal{E}(\rho)$, and gives:~\cite{schlosshauer,Hornberger:2003,joos2013decoherence,Vacchini_2009}
\begin{align}
    \rho_\mathcal{S}(\boldsymbol{x},\boldsymbol{x}',t) 
    &\rightarrow
    \rho_\mathcal{S}(\boldsymbol{x},\boldsymbol{x}',0) \bra{\chi_{\boldsymbol{x}'}} \ket{\chi_{\boldsymbol{x}}}  \label{eq:ex-eq}
\end{align}
By finding the overlap of the environment's out-states (using the scattering matrix), we can find the decoherence.
The scattering matrix is determined by the interaction's cross-section and thus differs depending on the type of decohering interaction. 
Typically, the overlap between the states will be an exponentially decaying function, which, as a result, causes a decay of the off-diagonal elements in the density matrix (see eq.~\eqref{eq:ex-eq}, which is where the 'quantumness' is stored.

Solving the dynamics of the density matrix exactly is incredibly hard, and assumptions are made to approximate the dynamics. The most common approximation we take is the Born approximation~\cite{born1926quantenmechanik}, which assumes weak coupling between the system and environment, neglecting higher-order coupling terms. 
Another assumption is that the initial state is a product state of the environment and system, i.e., that they are uncorrelated~\cite{bloch1956,redfield1957theory}.
Furthermore, taking the Markov approximation further simplifies the density matrix expression~\cite{lindblad1976generators,gorini1976completely}.
The Markov approximation, also called the short-memory approximation, assumes that self-interactions (memory effects) within the environment are short-lived compared to the timescale of the system's dynamics, such that these effects are negligible~\footnote{Studying non-Markovian systems can be challenging and has become a research topic in itself, see e.g. ref.~\cite{nonmarkovian}, we keep to the Markovian assumption here.}. 

Based on these assumptions and assuming that there are many scattering events and that the environment's particles are isotropically distributed in space, one can derive the standard Born-Markov master equation in eq.~\eqref{eq:deco-master} that can be found throughout literature; see refs.~\cite{schlosshauer,Hornberger:2003,joos2013decoherence,Vacchini_2009}.
The dynamics of the system's density matrix are given by this master equation:
\begin{equation}\label{eq:deco-master}
\pdv{\rho_\mathcal{S}(\boldsymbol{x},\boldsymbol{x}',t)}{t} = - \Gamma(\boldsymbol{x}-\boldsymbol{x}') \rho_\mathcal{S}(\boldsymbol{x},\boldsymbol{x}',t)
\end{equation}
where,
\begin{equation}\label{eq:deco-rate-def}
\begin{aligned}
    \Gamma(\boldsymbol{x}-\boldsymbol{x}') &= \int \dd{q} \rho(q) v(q) \int \frac{\dd{\boldsymbol{n}}\dd{\boldsymbol{n}'}}{4\pi} \abs{f(q\hat{\boldsymbol{n}},q\hat{\boldsymbol{n}}')}^2 \\ &\qq{}\left( 1 - e^{i q(\hat{\boldsymbol{n}}- \hat{\boldsymbol{n}}')\cdot (\boldsymbol{x}-\boldsymbol{x}')/\hbar} \right)
\end{aligned}
\end{equation}
is determined by the characteristics of the S-matrix via the cross-section $f(q\hat{\boldsymbol{n}},q\hat{\boldsymbol{n}}')$.
The $\rho(q)$ and $v(q)$ are the number density and velocity of environmental particles with momentum $q=\abs{\boldsymbol{q}}$, respectively. 
The unit vectors $\hat{\boldsymbol{n}}=\boldsymbol{q}/q$ and $\hat{\boldsymbol{n}}'=\boldsymbol{q}'/q'$ relate to the direction of the air particle's ingoing and outgoing momenta, and $\abs{f(q\hat{\boldsymbol{n}},q\hat{\boldsymbol{n}}')}^2$ is the differential cross section for a scattering $\boldsymbol{q}\to \boldsymbol{q}'$.
The factor $\Gamma(\boldsymbol{x}-\boldsymbol{x}')$ that characterizes the decoherence is often called the decoherence rate or localization rate.
Equation~\ref{eq:deco-rate-def} is quite common in decoherence literature~\cite{joos2013decoherence,schlosshauer}, and can also be derived from quantum field-theory perspective and using the operator-sum representation; see, for example, refs.~\cite{Kilian:2022kgm,Fragolino:2023agd}.

We can consider two limits to simplify the momentum integration that appears in eq.~\eqref{eq:deco-rate-def}: the limit where the wavelength of the environmental particle is much bigger than the spatial superposition (i.e. $\lambda_\text{air}\gg \Delta x$ and thus $q_0 \Delta x /\hbar \ll 1$, using the de Broglie relation), called the long-wavelength limit (LWL), and the limit where the wavelength of the environmental particle is much smaller than the spatial superposition (i.e. $\lambda_\text{air}\ll \Delta x$ and thus $q_0 \Delta x /\hbar \gg 1$ ), called the short-wavelength limit (SWL).
These limits are well known in the literature, see e.g.~\cite{schlosshauer}. 
In this work we compute eq.~\eqref{eq:deco-rate-def} for air molecules scattering without considering these approximations on the wavelength of the environmental particle.

\section{Wavelength-independent decoherence rate for Air molecule scattering}\label{sec:air-molecules}

The decoherence from scattering with air molecules has been found before both in the short-wavelength limit (SWL) and long-wavelength limit (LWL). 
Which limit is applicable depends on the superposition size $\Delta x$ (which is determined by the experiment) and on the thermal de Broglie wavelength of the air molecules, given by:
\begin{equation}\label{eq:thermal-wl-air}
    \lambda_\text{air} = \frac{2\pi\hbar}{\sqrt{2\pi m_\text{air} k_B T}} \, ,
\end{equation}
dependent on the external/ambient temperature.
The literature has considered different interpolations between the two wavelength regimes. 
In this section, we give an interpolation, show how it reproduces the SWL and LWL and compare our expression to the two wavelength limits and interpolations proposed in the literature.

\subsection{Decoherence expression}\label{subsec:exact-result}
To find the decoherence rate, we compute the integral in eq.~\eqref{eq:deco-rate-def}, which was given by:
\begin{equation}\label{eq:integral}
\begin{aligned}
    \int \dd{q} \rho(q) v(q) \int &\frac{\dd{\boldsymbol{n}}\dd{\boldsymbol{n}'}}{4\pi} \abs{f(q\hat{\boldsymbol{n}},q\hat{\boldsymbol{n}}')}^2 \\ &\times\big[ \underbrace{1}_\text{SWL-term} - \,\,\,\,\,\underbrace{e^{i q(\hat{\boldsymbol{n}}- \hat{\boldsymbol{n}}')\cdot (\boldsymbol{x}-\boldsymbol{x}')/\hbar}}_\text{LWL-term} \big]\, .
\end{aligned}
\end{equation}
Taking the wavelength limits often amounts to deciding which of the two terms dominates, as indicated in the equation above. We assume a geometrical cross section for the air molecule scattering with the particle of radius $R$~\footnote{%
The geometrical cross section is an assumption for the total cross section. 
Calculating collision cross section is a research field of its own, often based on numerical computations via specific methods such as DHSS and DTM and dependent on the modelling of the spheres and trajectory of the air molecules~\cite{larriba2013free}; this is outside the scope of this work.%
}:
\begin{equation}\label{eq:swlpart}
    \int \int \frac{\dd{\boldsymbol{\hat{n}}}\dd{\boldsymbol{\hat{n}}'}}{4\pi} \abs{f(q\boldsymbol{\hat{n}},q\boldsymbol{\hat{n}}')}^2 = \pi R^2\,.
\end{equation}
This cross-section gives the result of the integration over the solid angle for the first term between brackets in equation eq.~\eqref{eq:integral}, which is the term that dominates the decoherence rate in SWL. 

To perform the solid angle integral over the second term in between brackets in eq.~\eqref{eq:integral} (the part that dominates in the LWL), we consider:
\begin{align}
    \hat{\boldsymbol{n}} &= (\sin(\theta)\cos(\phi), \sin(\theta)\sin(\phi), \cos(\theta)) \, , \label{eq:n-vector-def} \\
    \dd{\boldsymbol{\hat{n}}} &= \sin{\theta}\dd{\theta}\dd{\phi}\, , \label{eq:dn-def}
\end{align}
and similarly for $\hat{\boldsymbol{n}}'$ with the angles $\theta'$, $\phi'$.
The integral over the exponent can be simplified by considering $\boldsymbol{x}-\boldsymbol{x}'$, which is generically a 3-dimensional vector, to be only in the $\boldsymbol{x}$-direction~\footnote{Here, we have assumed the superposition to be in the $\boldsymbol{x}$-direction, noting that other choices (i.e. the $\boldsymbol{y}$- or $\boldsymbol{z}$-direction) do not alter the final result. 
Ref.~\cite{schlosshauer}, on the other hand, averaged over the three directions in the dot product.
Interestingly, the resulting LWLs match despite the different approaches.}. 
With this simplification, the solid angle integral of the exponent-part of eq.~\eqref{eq:deco-rate-def} (i.e. the part indicated to be the `LWL-part') can be performed:
\begin{align}
    \int &\frac{\dd{\theta}\dd{\phi
    }\dd{\theta'}\dd{\phi'}}{4\pi} \sin(\theta) \sin(\theta') \abs{f(q\hat{\boldsymbol{n}},q\hat{\boldsymbol{n}}')}^2 \nonumber\\ &\qq{} e^{i q \Delta x [\sin(\theta)\cos(\phi) - \sin(\theta')\cos(\phi') ]/\hbar} \, . \\
    &=  \frac{\hbar^2 \pi R^2}{q^2 \Delta x^2} \sin^2\left(\frac{q\Delta x}{\hbar}\right) \, . \label{eq:lwlpart}
\end{align}
We recall that $q$ is the momentum magnitude of the environmental particle and the cross-section is given by the geometrical cross section.
After the solid angle integration, eq.~\eqref{eq:integral} (using eq.~\eqref{eq:swlpart} and eq.\eqref{eq:lwlpart}) becomes:
\begin{align}
    \int \dd{q} \rho(q) \frac{q}{m} \left( \pi R^2 - \frac{\hbar^2 \pi R^2}{q^2 \Delta x^2} \sin^2\left(\frac{q\Delta x}{\hbar}\right) \right) \, . \label{eq:mom-int}
\end{align}
The integration over the momentum magnitude remains, for which we consider $\rho(q)$ to be a Maxwell-Boltzmann distribution. The integration over momentum magnitude then gives:
\begin{equation}\label{eq:deco-exact}
\begin{aligned}
    \Gamma_\text{NWL}^\text{air} = &\frac{\sqrt{2 \pi} R^2 n_v }{m_\text{air}\sqrt{k_b T m_\text{air}}\Delta x} \bigg[2 k_b T m_\text{air} \Delta x  \\ &
    - \hbar \sqrt{2 k_b T m_\text{air}} \, \text{D}\left(\frac{\Delta x \sqrt{2 k_b T m_\text{air}}}{\hbar}\right)\bigg] \, . 
\end{aligned}
\end{equation}
This is the decoherence rate, which is generally denoted as $\Gamma$, the superscript `air' denotes that it is the decoherence due to air molecule scattering and the subscript `NWL' denotes that it does not use a wavelength approximation).
The function $\text{D}(x)$ is the Dawson function, also called the Dawson integral~\cite{abramowitz1968handbook,mccabe1974continued,weideman1994computation}. The Dawson function in eq.~\eqref{eq:deco-exact}
has two series expansions:
\begin{align}
    D(x) &\approx x - \frac{2}{3} x^3 + \frac{4}{15}x^5 \ldots \qq{} && \text{for } \abs{x}\ll1 \, , \label{eq:dfunc1}\\
    D(x) &\approx \frac{1}{2x} - \frac{1}{4x^3} + \frac{3}{8x^5} \ldots \qq{} && \text{for } \abs{x}\gg1 \, . \label{eq:dfunc2}
\end{align}
The argument of the Dawson function in eq.~\eqref{eq:deco-exact} is $2 \pi \Delta x/\lambda_\text{air}$ (using the expression for the wavelength given in eq.~\eqref{eq:thermal-wl-air}).
Thus, the two expansions correspond to the long and short wavelength limit, which we will discuss below.


\subsection{Long and short wavelength limits}\label{subsec:LWL}
In the long-wavelength limit (LWL), the wavelength of the ambient particle is considered to be much longer than the size of the superposition.
Usually, this limit is used to simplify the momentum integration in eq.~\eqref{eq:deco-rate-def}.
Since in this limit, $q_0 \Delta x /\hbar \ll 1$, the argument of the exponent in eq.~\eqref{eq:deco-rate-def} is small, and the exponent can be approximated by a Taylor expansion around zero. Following this prescription, various authors obtained the long wavelength limit, see ref.~\cite{schlosshauer}.

Regarding air molecule scattering, the LWL is applicable at higher temperatures when $\lambda_\text{air}>\Delta x$. 
By taking the appropriate limit: when the argument of the Dawson function $|x|\ll 1$, we get:
\begin{equation}\label{eq:deco-lwl}
    \Gamma_\text{LWL}^\text{air} = \frac{8 n_v R^2 }{3\hbar^2} \sqrt{2\pi m_\text{air}} (k_B T)^{3/2} (\Delta x)^2 
\end{equation}
which matches the LWL found via the approximation of the integral in eq.~\eqref{eq:deco-rate-def} in e.g. ref.~\cite{schlosshauer}.
In other words: 
$
    \Gamma_\text{ER}^\text{air} \overset{\Delta x \ll \lambda}{\approx} \Gamma_\text{LWL}^\text{air}
$.\newline


In the short-wavelength limit (SWL), the wavelength of the ambient particles is small compared to the superposition size. Again, this limit approximates the momentum integration of eq.~\eqref{eq:deco-rate-def} in a specific regime.
Since in this limit $q_0 \Delta x /\hbar \gg 1$, the exponential in eq.~\eqref{eq:deco-rate-def} oscillates rapidly and approximately averages out under integration. 
The exponential term is therefore neglected in this limit, and one considers only the geometrical cross-section ($\sigma_\text{tot}=\pi R^2$). 
By taking the appropriate limit of the argument of the Dawson function, i.e. $|x|\gg 1$, we get:
\begin{equation}\label{eq:deco-swl}
    \Gamma_\text{SWL}^\text{air}
    = 2 n_v R^2 \sqrt{\frac{2\pi k_B T}{m_\text{air}}} \, ,
\end{equation}
neglecting higher order $\lambda_\text{air}/\Delta x$-terms.
This matches our expression for the SWL found via the conventional method of approximating the integral in eq.~\eqref{eq:deco-rate-def}~\footnote{Note that eq.~\eqref{eq:deco-swl} differs a factor $8\pi/3$ from the SWL found in ref.~\cite{Biswas:2022qto}.}. In other words:
$\Gamma_\text{ER}^\text{air} \overset{\Delta x \gg \lambda}{\approx} \Gamma_\text{SWL}^\text{air}$.

Figure~\ref{fig:comparison} shows the result in eq.~\eqref{eq:deco-exact} and how it produces the SWL and LWL given in eqs.~\eqref{eq:deco-swl} and~\eqref{eq:deco-lwl}. 

\subsection{Comparison to literature}

\begin{figure}[bpt!]
    \centering
    \includegraphics[width=\linewidth]{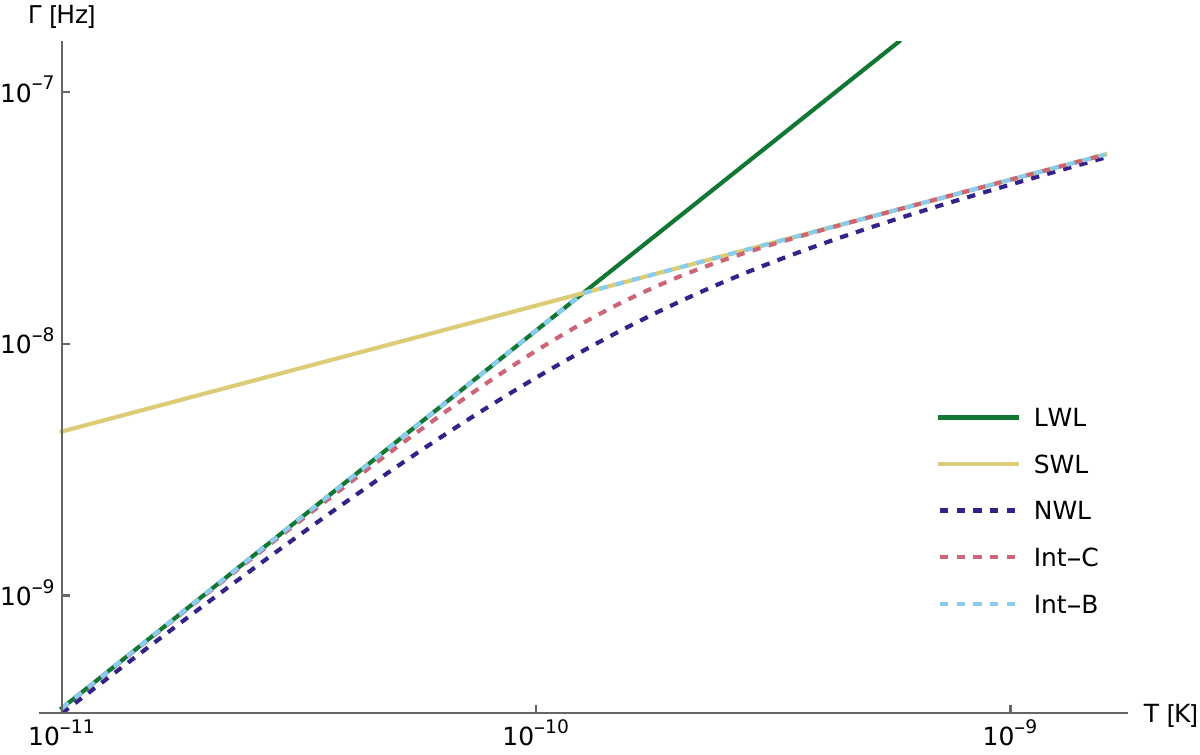}
    \caption{Comparison of the decoherence results from eq.~\eqref{eq:deco-exact} (the dashed darkpurple line labeled \textit{NWL}) to the interpolations used in~\cite{Biswas:2022qto} by Biswas et al. (the dashed light blue line labeled \textit{Int-B}, eq.~\eqref{eq:deco-biswas}) and the interpolation by Carlesso et al.~\cite{carlesso2016decoherence} (the dashed pink line labeled \textit{Int-C}, eq.~\eqref{eq:deco-carlesso}). 
    The modified long-wavelength limit, \textit{M-LWL} in eq.~\eqref{eq:deco-mlwl}) is not shown. The long-wavelength limit (LWL, solid green line, eq.~\eqref{eq:deco-lwl}) and short-wavelength limit (SWL, solid yellow line, eq.~\eqref{eq:deco-swl}) are also shown to match the three interpolations in the appropriate domains.}
    \label{fig:comparison}
\end{figure}

Besides the SWL and LWL limit, fig.~\ref{fig:comparison} also shows the comparison to the interpolations proposed in refs.~\cite{Biswas:2022qto} and~\cite{carlesso2016decoherence}.
One can interpolate between the SWL and LWL via several methods, and below, we briefly summarize the interpolations suggested by these two references, as well as an alternative approximation proposed in~\cite{Romero_Isart_2011}:
\begin{itemize}
\item {\bf Maximum and minimum interpolation}:
In ref.~\cite{Biswas:2022qto}, Biswas et al. interpolated between the two wavelength regimes by taking the minimal value of the two limits.
The decoherence function used for interpolation was:
\begin{equation}\label{eq:deco-biswas}
    \Gamma_\text{Int-B}^\text{air} = \text{min}(\Gamma_\text{SWL}^\text{air},\Gamma_\text{LWL}^\text{air}) \, .
\end{equation}
We label this decoherence rate `Int-B', referring to the Interpolation proposed by Biswas et al.
This interpolation gives a sharp switch between the wavelength limits, as can be seen from fig.~\ref{fig:comparison}.
\item {\bf Smooth interpolation}: Another possible interpolation was presented in ref.~\cite{carlesso2016decoherence} by Carlesso et al.:
\begin{equation}\label{eq:deco-carlesso}
    \Gamma_\text{Int-C}^\text{air} = \Gamma_\text{SWL}^\text{air} \tanh(\frac{\Gamma_\text{LWL}^\text{air}}{\Gamma_\text{SWL}^\text{air}}) \, .
\end{equation}
Using the hyperbolic tangent function gives a smoother transition between the wavelength regimes, as can be seen in fig.~\ref{fig:comparison}.
We label this decoherence function `Int-C', referring to the Interpolation proposed by Carlesso et al.
%
\item {\bf Saturated SWL}: When the physical situation is such that $\lambda_\text{air}<\Delta x$, i.e. the SWL is applicable, using the long-wavelength limit overestimates the decoherence rate.
This can be understood intuitively from the approximations and was shown in more detail in ref.~\cite{Fragolino:2023agd}.
In ref.~\cite{Romero_Isart_2011} Romero-Isart showed that taking the wavelength instead of the superposition width as the resolved length scale decreases this overestimation of the decoherence. 
They give the decoherence as $\Gamma = \Lambda \lambda_\text{air}^2$ (rather than the usual $\Gamma = \Lambda \Delta x^2$ in the LWL) to account for the saturation effect in the SWL limit.
Replacing $\Delta x$ by $\lambda_\text{air}$ in eq.~\eqref{eq:deco-lwl} gives:
\begin{align}
    \Gamma_\text{M-LWL}^\text{air} 
    &= \frac{8 n_v R^2 }{3\hbar^2} \sqrt{2\pi m_\text{air}} (k_B T)^{3/2} \lambda_\text{air}^2 \\
    &= \frac{16 \pi n_v R^2 }{3} \sqrt{\frac{2\pi k_B T}{m_\text{air}}} \label{eq:deco-mlwl}
\end{align}
This is not an interpolation but a modified LWL (as such, we label it `M-LWL') sometimes used in literature.
How well this decoherence function holds up depends on the superposition size, and we have not considered it in fig.~\ref{fig:comparison}, where it would give a line parallel to the SWL line.
\end{itemize} 

\begin{figure}[tpb!]
    \centering
    \includegraphics[width=\linewidth]{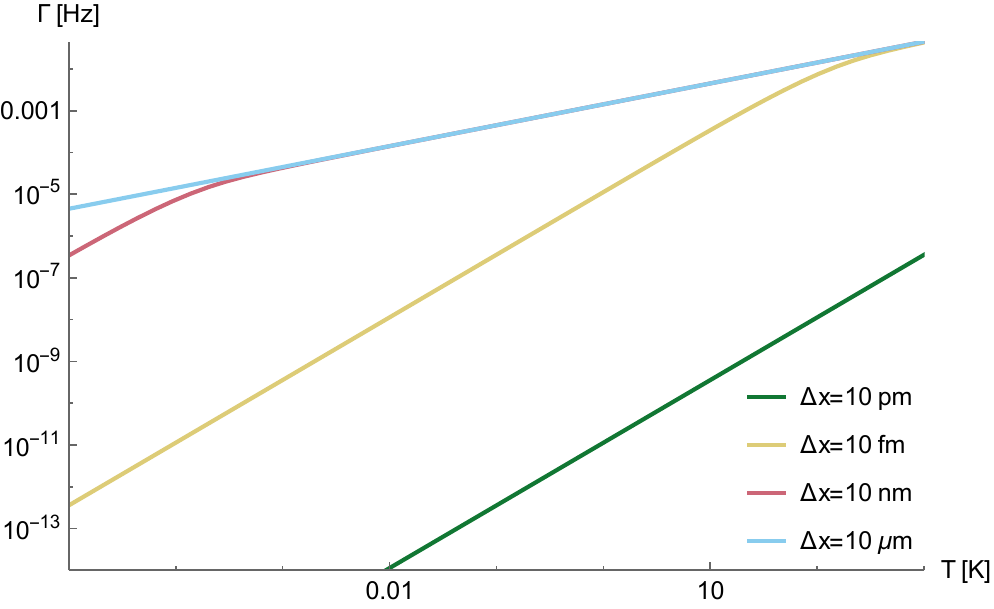}
    \caption{The decoherence rate based on eq.~\eqref{eq:deco-exact} for a sphere of radius $400\,\si{\nano\metre}$, for different superposition widths. 
    The lines corresponding to different superposition widths will overlap at higher temperatures due to the SWL being independent of the superposition width. At higher temperatures, the decoherence from blackbody scattering, absorption and emission (see footnote~\ref{footnote:blackbody}) could dominate the air molecule decoherence; the blackbody decoherence starts dominating earlier for larger superposition widths. The plot is made for $n_v = 10^8\,\si{\per\metre\cubed}$.}
    \label{fig:qgem1}
\end{figure}

Figure~\ref{fig:comparison} shows how the result in eq.~\eqref{eq:deco-exact} compares to these interpolations and how it reproduces the short-wavelength limit (SWL) and long-wavelength limit (LWL) in the appropriate domains.
Although in the figure a specific setup is chosen, namely $n_v=10^8\,\si{\per\metre\cubed}$, $\Delta x = 10^{-5}\,\si{\metre}$ and $R=0.4\,\si{\micro\metre}$, the relations between the lines shown in the figure hold in a more general sense.
The range for the experimental parameters and temperature range in Fig.~\ref{fig:comparison} are chosen to show the interpolation between the LWL and SWL domains; in sec.~\ref{sec:application} we discuss the application of the decoherence result in eq.~\eqref{eq:deco-exact} to a specific setup and focus more on showing the expected decoherence rates; see fig.~\ref{fig:qgem1}.

The M-LWL is not plotted since its validity depends entirely on the chosen superposition width; it becomes more appropriate for higher temperatures and/or larger superposition widths.
For the plot parameters, the M-LWL decoherence limit would be approximately parallel but one order of magnitude higher than the SWL line, it becomes a better approximation than the LWL at temperatures above nano-Kelvin.

\begin{table}[tpb!]
\captionsetup{justification=centering}
\setlength{\tabcolsep}{8pt} 
\renewcommand{\arraystretch}{1.5}
\captionsetup{skip=0pt}
\begin{center}
\begin{tabular}{||c|c|c c c||} 
 \hline
 $\Gamma$ & $T$ & $\Delta x$ & $n_v \, (\si{\per\metre\cubed})$  & $P \, (\si{\pascal})$ \\ [0.5ex] 
 \hline\hline
 \multirow{4}{*}{$0.01\,\si{\hertz}$}
&\multirow{4}{*}{$1\,\si{\milli\kelvin}$}
 & $10 \, \si{\femto\metre}$ & $3*10^{21}$ & $3.9*10^{-5}$ \\
 \cline{3-5} & & $10 \, \si{\pico\metre}$  & $3*10^{15}$ & $3.9*10^{-11}$ \\
 \cline{3-5} & & $10 \, \si{\nano\metre}$  & $2*10^{10}$ & $3.2*10^{-16}$ \\
 \cline{3-5} & & $10 \, \si{\micro\metre}$  & $2*10^{10}$ & $3.1*10^{-16}$ \\ 
 \hline
 \multirow{4}{*}{$0.05\,\si{\hertz}$}
 &\multirow{4}{*}{$1\,\si{\kelvin}$} 
 & $10 \, \si{\femto\metre}$ & $4*10^{17}$ & $6.1*10^{-6}$ \\
 \cline{3-5} & & $10 \, \si{\pico\metre}$  & $4*10^{11}$ & $6.1*10^{-12}$ \\ 
 \cline{3-5} & & $10 \, \si{\nano\metre}$  & $4*10^9$ & $4.9*10^{-14}$ \\ 
 \cline{3-5} & & $10 \, \si{\micro\metre}$  & $4*10^9$ & $4.9*10^{-14}$ \\ 
 \cline{2-5}
 \hline\hline
  \multirow{12}{*}{$0.1\,\si{\hertz}$}
&\multirow{4}{*}{$1\,\si{\milli\kelvin}$}
 & $10 \, \si{\femto\metre}$ & $3*10^{22}$ & $3.9*10^{-4}$ \\
 \cline{3-5} & & $10 \, \si{\pico\metre}$  & $3*10^{16}$ & $3.9*10^{-10}$ \\
 \cline{3-5} & & $10 \, \si{\nano\metre}$  & $2*10^{11}$ & $3.2*10^{-15}$ \\
 \cline{3-5} & & $10 \, \si{\micro\metre}$  & $2*10^{11}$ & $3.1*10^{-15}$ \\ 
 \cline{2-5}
 &\multirow{4}{*}{$1\,\si{\kelvin}$} 
 & $10 \, \si{\femto\metre}$ & $9*10^{17}$ & $1.2*10^{-5}$ \\
 \cline{3-5} & & $10 \, \si{\pico\metre}$  & $9*10^{11}$ & $1.2*10^{-11}$ \\ 
 \cline{3-5} & & $10 \, \si{\nano\metre}$  & $7*10^9$ & $9.7*10^{-14}$ \\ 
 \cline{3-5} & & $10 \, \si{\micro\metre}$  & $7*10^9$ & $9.7*10^{-14}$ \\ 
 \cline{2-5}
 &\multirow{4}{*}{$4\,\si{\kelvin}$}
 & $10 \, \si{\femto\metre}$ & $1*10^{17}$ & $6.1*10^{-6}$ \\
 \cline{3-5} & & $10 \, \si{\pico\metre}$  & $1*10^{11}$ & $6.2*10^{-12}$ \\
 \cline{3-5} & & $10 \, \si{\nano\metre}$  & $4*10^9$ & $1.9*10^{-13}$ \\
 \cline{3-5} & & $10 \, \si{\micro\metre}$  & $4*10^9$ & $1.9*10^{-13}$ \\
 \hline
\end{tabular}
\end{center}
\caption{\justifying Experimental parameters needed for a certain decoherence rate $\Gamma$ from air molecule scattering, for different superposition sizes $\Delta x$.
The experimental parameters are the ambient temperature $T$, the pressure $P$ and the number density $n_v$, obtained via the ideal gas law.
The decoherence rate, given in eq.~\eqref{eq:deco-exact}, scales linearly with the number density, as seen in the table.
The table is given for a microsphere of radius $400\,\si{\nano\metre}$ and air molecules of $28.97 \text{a.m.u}$.
At higher temperatures ($>4\,\si{\kelvin}$) blackbody radiation becomes a considerable decoherence channel as well, see footnote~\ref{footnote:blackbody}.}
\label{table:parameters}
\end{table}

\section{Application: Mesoscopic spatial superpositions}\label{sec:application}

The result for the decoherence rate via air molecule scattering was given in eq.~\eqref{eq:deco-exact} and is now applied to a specific experimental setup.
We consider the creation of a spatial superposition in a nano/microsphere in a matter-wave interferometer, following the experimental protocol described in ref.~\cite{Bose:2017nin}, called the QGEM proposal.
The QGEM protocol is based on bringing two Stern-Gerlach interferometers (with diamond test masses with embedded spins) adjacent, such that the gravitational interaction can entangle the two superpositions~\cite{Bose:2017nin,ICTS}.
If gravity is inherently quantum, it would entangle the two masses of the matter-wave interferometers and thus certify the quantum nature of gravity via witnessing the gravitationally-induced entanglement in the spin correlations~\cite{Chevalier:2020uvv}.
The theoretical descriptions were elaborated in Refs.~\cite{Marshman:2019sne,Bose:2022uxe,Vinckers:2023grv}.
Since the protocol considers nano/microspheres, it can suffer greatly from decoherence. This section analyses the experimental parameters needed for a given coherence time based on decoherence from air molecule scattering.
In sec.~\ref{subsec:time-dependence}, we also consider more carefully the time-dependence of the decoherence rate in a Stern-Gerlach interferometer.
In such an interferometer, the embedded spin creates and annihilates the spatial superposition; since the superposition width changes during the experiment, the decoherence rate becomes time-dependent.

Figure~\ref{fig:qgem1} shows the decoherence rate as a function of temperature for different superposition sizes.
Although for the witnessing of gravitationally induced entanglement, one needs at least micrometer-size spatial superposition, other entanglement experiments could be done with smaller superposition sizes.
Additionally, the experimental realization of mesoscopic spatial superpositions will start with small superposition sizes and then work towards enlarging this.
Therefore, we consider a range of superposition sizes in fig.~\ref{fig:qgem1}.
The figure shows that the lines corresponding to the $10$ nm and $10\,\si{\micro\metre}$ superposition width overlap at higher temperatures. 
This is because the short-wavelength limit is independent of the superposition width; the SWL gives the decoherence rate as temperature increases.
This also happens for the smaller superposition widths at higher temperatures, which is visible for the femtometer size line, but outside the temperature range for the picometer one.
Figure~\ref{fig:qgem1} is plotted for $n_v = 10^8\,\si{\per\metre\cubed}$, meaning $P\in[10^{-20},10^{-12}]\,\si{\pascal}$.

In Table.~\ref{table:parameters}, we have presented an overview of the necessary experimental parameters for a coherence time $1/\Gamma$ based on the air molecule decoherence.
Although the final goal in the context of the QGEM protocol will be to create micron-size superpositions for testing gravity-induced entanglement (expected to be characterized by the entanglement rate $\omega_\text{ent}\approx 0.01-0.1\,\si{\hertz}$ for diamond nano/microcrystals), a first step will be to create femtometer-size superpositions and we, therefore, give the experimental parameters for a range of superposition sizes for decoherence rates $0.01 - 0.1\,\si{\hertz}$.
The radius of the microparticle for the table values is taken to be $0.4\,\si{\micro\metre}$, which in the case of diamond (the typical material considered in the QGEM protocol) would correspond to $10^{-15}\,\si{\kilogram}$; the mass of air molecules is taken to be $28.97$ u.

The results in Tab.~\ref{table:parameters} are repeated for different temperatures, showing that a higher temperature requires for a smaller pressure. 
Especially for the small superposition sizes, these experimental parameters should be realizable with current technologies.
It should be noted that at temperatures above $\sim 4\,\si{\kelvin}$, blackbody radiation becomes a dominant source of decoherence for micrometer-size superpositions in, e.g., diamond~\footnote{
Unfortunately, the blackbody case does not have a straightforward interpolation (additionally, numerical computation of the generic solution is problematic). However, we can still use the two wavelength limits.
For temperatures $T\in[\si{\micro\kelvin},300\si{\kelvin}]$ the wavelength of light is in $\lambda_\text{th}\in[40\si{\micro\meter},10^{10}\si{\micro\metre}]$. Thus, the blackbody can almost always be considered in the LWL.
The decoherence rate in the long-wavelength limit has been known in literature~\cite{schlosshauer,Sinha:2022snc,Romero_Isart_2011,Biswas:2022qto} for the emission, absorption and scattering of blackbody photons.\label{footnote:blackbody}}. 
The table also shows the experimental parameters for different decoherence rates; since this rate scales linearly with temperature and pressure, we have not given an exhaustive overview of many decoherence rates.
Instead, the experimental parameters for other decoherence rates can easily be deduced from the values shown in the table.

\subsection{Time-dependent Mesoscopic  interferometer}\label{subsec:time-dependence}
In general, superpositions realized in Stern-Gerlach interferometers are time-dependent, see~\cite{
PhysRevLett.125.023602,Marshman:2021wyk,Zhou:2022epb,Zhou:2022frl,Zhou:2022jug,Zhou:2024voj,Braccini:2024fey}. 
Therefore, the decoherence rate of the system will also be time-dependent. 
We encode this time dependency directly in eq.~\eqref{eq:deco-master}, since it affects the superposition distance. 
The differential equation becomes:
\begin{align}\label{eq:deco-master-time-dependent}
\pdv{\rho(\boldsymbol{x},\boldsymbol{x}',t)}{t} = - \Gamma(\boldsymbol{x}-\boldsymbol{x}',t) \rho(\boldsymbol{x},\boldsymbol{x}',t)\,,
\end{align}
for $\boldsymbol{x}\neq \boldsymbol{x}'$ (the decoherence does not appear on the diagonal terms of the density matrix). 
Given a time-dependent parametrization, $\Delta x(t)$, for the superposition distance, we can use the decoherence rate found in sec.~\ref{sec:air-molecules} and apply it in the time-dependent case.
Assuming a Stern-Gerlach interferometer, where a spin qubit in a magnetic field gradient is used to create a spatial qubits, the spin state $\ket{\uparrow}$, $\ket{\downarrow}$ is coupled to the position state $\ket{\boldsymbol{x}}$, $\ket{\boldsymbol{x}'}$, respectively.
The solution for eq.~\eqref{eq:deco-master-time-dependent} is given by:
\begin{equation}\label{eq:time-dependent-solution}
\begin{aligned}
\rho(\tau) = \frac{1}{2} \bigg[ &\ket{\uparrow,\boldsymbol{x}}\bra{\uparrow,\boldsymbol{x}} + \ket{\downarrow,\boldsymbol{x}'}\bra{\downarrow,\boldsymbol{x}'} \\ &+ \ket{\uparrow,\boldsymbol{x}}\bra{\downarrow,\boldsymbol{x}'} e^{-\Gamma(\tau)} + \ket{\downarrow,\boldsymbol{x}'}\bra{\uparrow,\boldsymbol{x}} e^{-\Gamma(\tau)}\bigg] \,,
\end{aligned}
\end{equation}
where the decoherence rate $\Gamma$ is time-dependent and given by $\Gamma(\tau) = \int\limits_{0}^{\tau} \dd{t} \Gamma(\boldsymbol{x}-\boldsymbol{x}',t)$, with $\tau$ the experimental time and the function $F$ as in eq.~\eqref{eq:deco-rate-def} but with a time-dependent superposition size width given by $\Delta x(t) = \vert \boldsymbol{x}(t) - \boldsymbol{x}'(t)\vert$.
\begin{figure}[bpt!]
    \centering
    \includegraphics[width=1.375\linewidth]{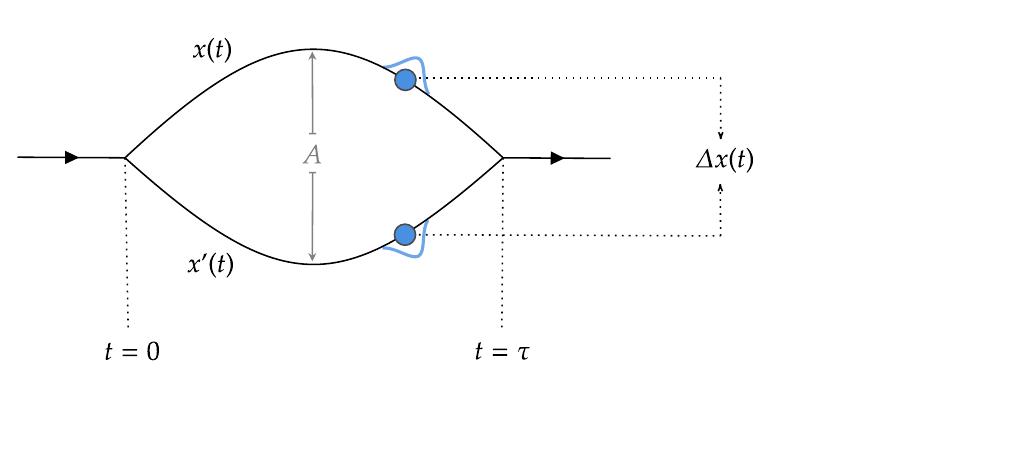}
    \caption{\justifying
    Example of time-dependent trajectories for creating the superposition. 
    Each superposition path follows a sine-like path with amplitude $A/2$, such that the maximal separation is $A$. 
    The parametrization is constructed so that the superposition initializes at time $t=0$ and ends at $t = \tau$.}
    \label{fig:sine-path}
\end{figure}

The parameterization of $\Delta x(t)$ depends on the scheme used to create the superpositions, for example, we choose a simple path to show the relevance of the time-dependent calculation. 
Our parametrization, as illustrated in fig.~\ref{fig:sine-path}, has superposition paths $x(t)$ and $x'(t)$ given by:
\begin{align}
    x(t) = \frac{A}{2}\sin\left(\frac{\pi t}{\tau}\right) \ , \ \ x'(t) = -\frac{A}{2}\sin\left(\frac{\pi t}{\tau}\right) \, .
\end{align}
Our naive example of the sine function is used solely to illustrate the point that the decoherence rate is time-dependent.
The size of the spatial superposition during the experimental time $t\in[0,\tau]$ is given by:
\begin{align}
    \Delta x(t) = A \sin\left(\frac{\pi t}{\tau}\right).
\end{align}
We numerically compare the time-dependent and constant decoherence rates. For the time-dependent superposition, we consider the trajectory shown in fig.~\ref{fig:sine-path} with $\Delta x(0) = 0$ at the initialization of the superposition, reaching a maximal separation of size $A$ at $t=\tau/2$, and vanishing again at the instant $t = \tau$.
For comparison, we consider the constant decoherence with fixed superposition size at $\Delta x = A$.

There are many different schemes for creating spatial superpositions in mesoscopic test masses~\cite{
PhysRevLett.125.023602,Marshman:2023nkh,Zhou:2022epb,Zhou:2022frl,Zhou:2022jug,Zhou:2024voj,Braccini:2024fey}, each of which giving a different parametrization for $\Delta x(t)$.
Currently, these works on mesoscopic spatial superpositions are theoretical proposals that have not been experimentally realized. 
For example, ref.~\cite{Margalit:2020qcy} proposed a trajectory for a Stern-Gerlach interferometer, later fine-tuned by ref.~\cite{Marshman:2021wyk}, which is approximately sine-like.
These proposed protocols~\cite{Margalit:2020qcy,Marshman:2021wyk} are approximated by the simple trajectory in fig.~\ref{fig:sine-path}.
Reference~\cite{Margalit:2020qcy} showed experimentally, as proof of principle, that their scheme can be used to create large ($\sim1\,\si{\micro\metre}$) superposition in a Bose-Einstein condensate of $10^4$$~^{87}\text{Rb}$ atoms.

\begin{figure}[tpb!]
    \centering
    \includegraphics[width=\linewidth]{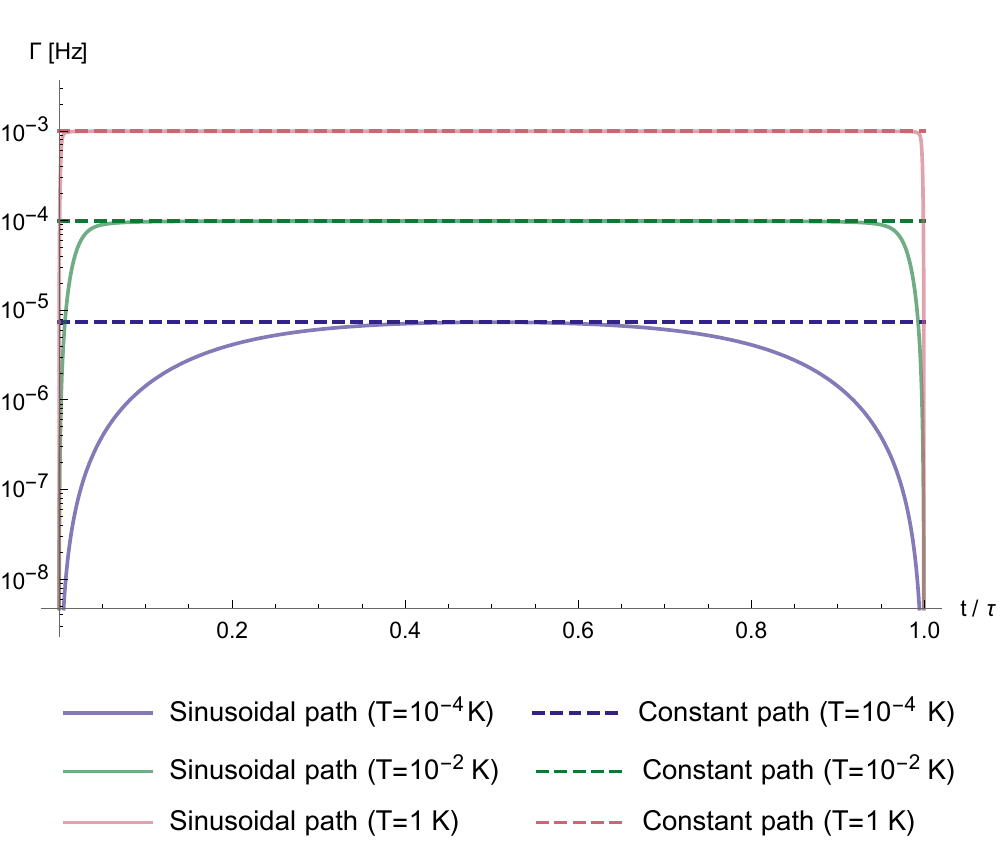}
    \caption{\justifying
     Comparison between decoherence rates (from eq.~\eqref{eq:deco-exact}) for superpositions that interact with gas particles under different temperatures, following the sinusoidal path (full lines) shown in figure \ref{fig:sine-path}, and a path with constant separation $A$ (dashed lines). $\tau$ denotes the total experimental time, and different colours represent different temperatures: the red, orange and blue lines correspond to ambient temperatures of, respectively, $T=1$ K, $T=10^{-2}$ K and $T=10^{-4}$ K for $\Delta x =10{\rm nm}$ superposition size. 
     The other parameters used correspond to the maximal separation of $A=10$ nm and the number density of particles $n_v=10^{8}$ m$^{-3}$. The radius of the scatterer is considered to approximately $R=0.4\,\si{\micro\metre}$.}
    \label{fig:path-comparison}
\end{figure}

Figure~\ref{fig:path-comparison} shows the decoherence rate given in eq.~\eqref{eq:deco-exact} compared to the constant-path approach, as a function of time.
The figure shows that the constant-path approach, usually assumed for estimating decoherence in spatial superposition, can overestimate the effect of decoherence.
One should note that only the LWL and our interpolation display time dependence since they are dependent on the superposition size; see eq.~\eqref{eq:deco-lwl}.
The SWL in eq.~\eqref{eq:deco-swl}, on the other hand, is independent of $\Delta x$~\footnote{The SWL could contain time-dependent corrections if one considers varying cross-sections, which are beyond the scope of this work.} and its decoherence rate is not time-dependent.
Therefore, differences between the constant-superposition-size approach and the time-dependent one only appear in the LWL and when interpolating between the SWL and LWL limits.
The environmental temperatures corresponding to the different coloured lines in the fig.~\ref{fig:path-comparison} are chosen to illustrate this point.
In fig.~\ref{fig:path-comparison}, the three lines correspond to the different wavelength regimes: for the red line, the air molecule wavelength is much smaller than the maximal superposition size. Because this is the shortest wavelength situation, this also gives the most decoherence.
For the orange line in fig.~\ref{fig:path-comparison}, the wavelength is of the same order as the maximal superposition size, and for the blue line the wavelength is much larger than the maximal superposition size (thus giving the smallest decoherence of the three lines).
At the start of the experiment, we will always be in the LWL, since the superposition size starts at zero. 
Later, we can enter the interpolation regime between the two wavelength limits, which is the case for the orange and red lines in fig.~\ref{fig:path-comparison}. 
The figure illustrates nicely that it is temperature-dependent when the interpolation regime is entered, i.e. we are no longer completely in the LWL.
For higher ambient temperatures, the transition between the wavelength regimes happens quickly, since the ambient particles have a relatively short wavelength; the constant-path approach is a good approximation of the decoherence rate.
%

Given a superposition width of maximally $10\,\si{\nano\metre}$, fig.~\ref{fig:path-comparison} shows that for a temperature of $\sim 1\,\si{\kelvin}$ and higher, even when taking into account the creation/annihilation of the superpositions, the constant-path approach is a good approximation for the decoherence rate~\footnote{Of course, the figure will be slightly different for different paths, i.e. different schemes for creating the superposition, but as long as the maximum superposition size is the same and superposition is created quickly, this statement should hold.}.
The constant-path approximation overestimates the decoherence rate for $T \leq 1$ K for $10\,\si{\nano\metre}$ superpositions. 
Note that by increasing the superposition size $\Delta x$ (while keeping the other parameters above fixed), the curves in fig.~\ref{fig:path-comparison} (for the same temperatures) would flatten/saturate the SWL more quickly.
In other words, the system's coherence becomes increasingly fragile to scattering for arbitrarily large superposition sizes.\newline

\section{Conclusion}\label{sec:conclusion}

This work presents an expression for the decoherence rate due to air molecules scattering with spatial superpositions that does not use any wavelength limits. 
Previously, this decoherence rate was only computed in two regimes: long and short wavelength compared to the size of the superposition.
Our result reproduces the known SWL and LWL in the appropriate domains.
In fig.~\ref{fig:comparison}, we have shown that these limits are met, and we have compared the our interpolation result, presented in eq.~\eqref{eq:deco-exact}, to other interpolations found in previous works~\cite{carlesso2016decoherence,Biswas:2022qto}.

In sec.~\ref{sec:application}, we have given an overview of the necessary experimental parameters (i.e. ambient temperature and pressure), given a specific decoherence rate and for several spatial superposition sizes.
We hope the results in tab.~\ref{table:parameters} can guide the experimental parameters of mesoscopic spatial superpositions for the collisional scattering-induced decoherence rates.

Additionally, we showed that one should be careful not to overestimate the decoherence effects by neglecting the time dependence of the superposition size. Taking the superposition size to be constant in time works well as an approximation if one quickly enters the SWL.
This is, for example, the case for nanometer-size superpositions in mesoscopic spatial superpositions at $1\,\si{\kelvin}$.
We hope that these parameters and various limits will guide the matter-wave interferometer community towards realising mesoscopic spatial superpositions. 
In future work, we hope to also find an interpolation result for the decoherence of a mesoscopic object due to the blackbody scattering, emission and absorption. \newline

\section*{Acknowledgements}

S.B. and A.M's research is funded by the Gordon and Betty Moore Foundation through Grant GBMF12328, DOI 10.37807/GBMF12328, and the Alfred P. Sloan Foundation under Grant No. G-2023-21130. SB
thanks EPSRC grants EP/R029075/1, EP/X009467/1,
and ST/W006227/1.
\bibliography{casimir.bib} 
\bibliographystyle{ieeetr}

\end{document}